 \documentclass[preprint2]{aastex}

\newcommand{\feh}{\ensuremath{[\mbox{Fe}/\mbox{H}]}}
\newcommand{\rphk}{\ensuremath{R'_{\mbox{\scriptsize HK}}}}
\newcommand{\lrphk}{\ensuremath{\log{\rphk}}}
\newcommand{\caii}{\ion{Ca}{2} H \& K}
\newcommand{\mv}{\ensuremath{M_{\mbox{\scriptsize V}}}}
\newcommand{\ms}{\ensuremath{M_{\mbox{\scriptsize V,MS}}}}
\newcommand{\dms}{\ensuremath{\Delta \mv}}

\shorttitle{Maunder Minimum stars}
\shortauthors{Wright}

\begin{document}
\title{Do We Know of any Maunder Minimum Stars?\altaffilmark{1}}
\altaffiltext{1}{Based on observations obtained
at the W. M. Keck Observatory, which is operated jointly by the
University of California and the California Institute of Technology.
The Keck Observatory was made possible by the generous financial support of the W.M. Keck Foundation.}
\author{J. T. Wright}

\affil{Department of Astronomy, University of California at
Berkeley, 601 Campbell Hall, Berkeley, CA 94720-3411}
\email{jtwright@astro.berkeley.edu}

\begin{abstract}

Most stars previously identified as Maunder minimum stars are
old stars evolved off of the main sequence.  Analysis of
activity measurements from the California and Carnegie Planet Search
program stars and Hipparcos parallaxes implies that the
canonical age-chromospheric activity relation breaks down for stars older than
$\sim 6$ Gyr when activity is calculated from Mount Wilson S values.
Stars only 1 magnitude above the main sequence exhibit
significantly suppressed activity levels which have been mistaken for
examples of Maunder minimum behavior.  

\end{abstract}


\section{Introduction}

Studies of the star formation history of the galaxy, searches for
extrasolar planets, and studies of the Sun as a star often benefit from
the ability to determine the age of sun-like stars.  But, because
stars change very little during their lifetime on the  
main sequence, determining their age is difficult.  
An important tool for making that determination is the simple,
monotonic relation between age and chromospheric activity.  As stars
age, they lose angular momentum and have less magnetic 
activity on their surfaces.  Magnetic fields lead to chromospheric
heating, which is balanced by cooling in, among others, the \caii\
lines.  The strength of these emission lines is thus an indicator of stellar 
age.  Modulation of these lines reveals rotation (as active regions
rotate into and out of view) and stellar activity cycles.

In 1966 Olin Wilson began a program to measure the strength of the
\caii\ lines in 91 stars as a function of time.  This program has been
expanded over the years to include many more stars on larger
telescopes and better spectrometers \citep{Vaughan1978,Baliunas1995}.
Since then, many other projects have measured activity levels in main
sequence stars with different instruments and aims
\citep[e.g.][]{Henry1996,Strassmeier2000,Tinney2002}.

Today, the canonical age-chromospheric activity relation is that of
\citet{Soderblom1991}, as refined by \citet{Donahue1993} and \citet{BaliunasBAAS1995}:
\begin{equation}
  \label{ageact}
   \log{\mbox{Age}/\mbox{yr}} = -0.0522 R_5^3 + 0.4085 R_5^2 -1.334
   R_5 +10.725
\end{equation}
where $R_5 = \rphk \times 10^5$ and \rphk\ represents an
activity level for a star averaged over many stellar activity cycles.
\citet{Donahue1998} provides an excellent discussion of this
age-chromospheric activity relation, which describes a
one-to-one relationship between activity and age. 

Surveys of activity in sun-like stars provide an opportunity to get a
global picture of the sun's activity history
\citep[e.g.][]{Zhang1994, Lockwood1992}.  In particular, detecting states of
extremely low activity in young stars analogous to the Maunder
Minimum in the Sun has been of interest.  For instance, Maunder minima
have been examined as a probe of the nature
of the solar dynamo \citep{Saar1992}, and as they relate to changes in
solar luminosity and assumptions in climate models \citep{Baliunas1990}.

Activity in evolved stars, measured by the Mount Wilson project as
well as others \citep[e.g.][]{Strassmeier1994,Dupree1995},
is also well-studied.  \citet{do Nascimento2003} showed that
evolved stars with $\bv > 0.55$ have lower activity levels than their main
sequence counterparts, leading to the possibility of confusion between
slightly evolved stars and stars in a Maunder Minimum-like state.

\section{Data}
\subsection{Activity measurements\label{data}}

This paper discusses data from four sources:  the activity
catalogs of the California \& Carnegie Planet Search Program
\citep[][henceforth WMBV]{Wright2004}, the Mount Wilson program
\citep{Duncan1991,Vaughan1978}, an activity survey for Project Phoenix
\citep{Henry1996}, and the Hipparcos catalog
\citep{Hipparcos, Perryman1997}.  To increase the confidence in the results here,
only those WMBV stars with more than two Keck observations are used in
this work.

The bulk of the $\sim 1000$ stars in WMBV are currently being monitored for
radial velocity variations and are typically bright, within two
magnitudes of the main-sequence,
and of spectral type F7 or later.  The Mount Wilson project includes over
1000 stars of all spectral types and luminosity classes.  The Project
Phoenix catalog is the result of an activity survey of $\sim 800$ bright
southern stars originally selected as part of SETI.  This work uses the
combined data set of all of the samples in the Project Phoenix catalog.  The
Hipparcos catalog provides $M_V$ and \bv\ for nearly all of the stars
in the above catalogs. 

Activity measurements in main sequence stars are usually presented as
$S$-values or as \rphk-values.  $S$ is essentially the equivalent
width of the emission cores in the \caii\ lines.  Typical $S$-values
range with spectral type from $\sim 0.1$ to 1.0 or more (for dMe
stars).  \rphk\ is a measurement of the fraction of a star's total
luminosity emitted in the \caii\ line cores, excluding the  
photospheric component which would be present in the absence of any
chromospheric heating.  Here, I have transformed the Mount Wilson
catalog $S$-values to \rphk\ according to the same prescription used in
\citet{Henry1996} and WMBV.  A discussion of these quantities can be found
in any of the references to activity catalogs listed above.  Note that there
is an error in equation 10 of WMBV:  the left hand side should read
``$\log{C_{\mbox{cf}}}(\bv)$''.

\subsection{Errors\label{errors}}

Uncertainties in $M_V$ and \bv\ are documented in the Hipparcos
catalog.  Typical errors for the kinds of stars considered here are
0.001 \arcsec\ in parallax and 0.01 magnitude in \bv. 
Uncertainty in \rphk\ is more difficult to determine because 
many stars exhibit variable activity levels.  In WMBV, for instance,
the single-observation random error in the $S$ index is estimated to be
6\%, but comparison to Mount Wilson $S$ values reveals a typical difference of
13\%, probably because the measurements are not contemporaneous.  A
6\% uncertainty corresponds to a uncertainty of $0.025$ in \lrphk.

\section{Color-Magnitude Diagrams and the Main Sequence}
\subsection{Definition of the Main Sequence\label{ms}}
The Hipparcos catalog allows one to determine an analytic
expression for the locus in an H-R diagram of the average main
sequence in the solar neighborhood.  To determine this expression, I
have used only stars in the Hipparcos
catalog within 60 pc and with robust ($4\sigma$) parallaxes.

Plotting this sample on a color-magnitude diagram, I fit a straight line to the
main sequence by eye, then fit a second-order polynomial to those
points within 1.3 $V$ magnitudes of this line.  I then iteratively fit
progressively higher-order polynomials to those points within a
progressively smaller vertical distance of the previous fit.  This
rejection of points prevented evolved and pre-main sequence stars from
contaminating the final fit.

\begin{figure}
\plotone{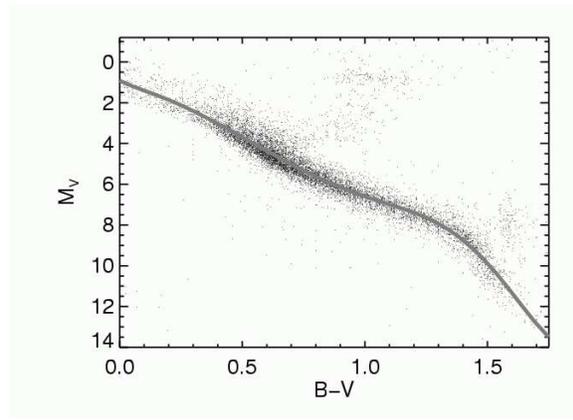}
\caption{Polynomial fit to the Hipparcos average main
  sequence.  Hipparcos stars within 60 pc were used for this fit, as
  described in Section~\ref{ms}.\label{msfig}}
\end{figure}

This algorithm produced an eighth-order polynomial expression for
$\mv$ for the main sequence, $\ms(\bv) = \sum a_i (\bv)^i$, with coefficients
$a = \left\{\right.$ 0.909, 6.258, -23.022, 125.5537, -321.1996, 485.5234,
-452.3198, 249.6461, -73.57645, 8.8240$\left.\right\}$.  This fit is shown in
Figure~\ref{msfig}.    

Then, I define, 
\begin{equation} 
\dms = \ms(\bv) - M_V 
\end{equation}  
\dms\ represents the height above the Hipparcos average main sequence
for a star with color \bv\ and absolute magnitude $M_V$.  With this definition,
evolved stars and metal-rich stars have positive values of 
\dms.   

\begin{figure}
\plotone{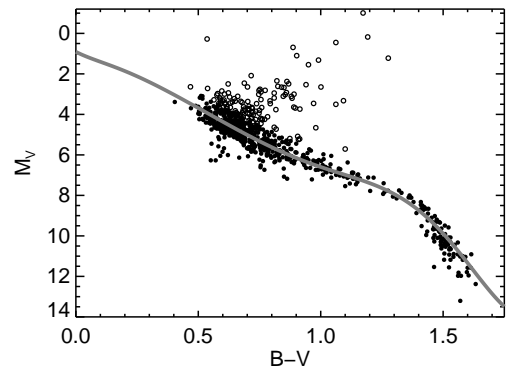}
\caption{Hipparcos average main sequence from Figure~\ref{msfig} and stars from the
California \& Carnegie Planet Search.\label{fefit}}
\end{figure}

Figure~\ref{fefit} shows the California \& Carnegie Planet Search
sample on a color-magnitude diagram with the Hipparcos average main
sequence.  Stars with $\dms > 1$ are plotted with open symbols.

\subsection{Effects of Evolution and Metallicity\label{evolution}}
There is a degeneracy between the effects of metallicity and evolution
in a star's position on a color-magnitude diagram:  both evolution and
enhanced metallicity place stars to the right of the main sequence.  Based upon
the theoretical isochrones of \citet{Girardi2002}, I estimate
that a star with $\feh = +0.3$ will lie $\sim 0.45$ magnitude above the
main sequence of a solar-metallicity star, and a star with $\feh=-0.3$
will lie $\sim 0.45$ magnitude beneath it.  

Stars evolve very slowly while they are on the main sequence, so \dms\
changes very little until a star begins to evolve toward the end of its
lifetime.  That means that the quantity \dms\ is insensitive to the age of main
sequence stars since movement on a color-magnitude diagram during the
main-sequence phase is small compared to changes in position with
metallicity.  Once a star is old enough to have
undergone significant changes in its physical structure, however,
\dms\ tracks age more sensitively.  As stars of similar mass age and
lose angular momentum, activity will decrease with \dms.  Stars of
different masses, however, will leave the main sequence at different
ages, so for heterogeneous samples of stars \dms\ is a poor proxy
for age without the benefit of accurate models and precise metallicities.

\section{Maunder Minima, Age, and Evolution}
\subsection{Maunder Minimum Stars}
Many surveys of stellar activity have noted the existence of a
population of stars with very low activity.  \citet{Baliunas1990}
noted that in the Mount Wilson survey some 30\% of sun-like stars
appeared to be in a states of low 
activity, dubbed ``Maunder minima'', analogous to the Maunder
Minimum (distinguished by a capital {\it M}) of the Sun in the 17th century.
\citet{Saar1992} revised this estimate to 10-15\% and other
authors have since reached similar conclusions with different 
data sets \citep{Henry1996,Baliunas1995,Saar1998,Gray2003}.  \citet{Henry1996}
divided their sample into four activity categories from ``very
active'' stars, with $\lrphk > -4.2$ to ``very inactive'' stars with
$\lrphk < -5.1$, the latter of which they considered Maunder minimum
candidates. 

\citet{Baliunas1990} suggested that the activity level of the sun
during the Maunder Minimum corresponded to that of these 
low-activity stars, $S=0.145$\footnote{$S=0.145$ for the Sun corresponds to
$\lrphk = -5.1$.  During a typical solar minimum, $S_\odot = 0.165$,
corresponding to $\lrphk=-4.97$.}.  \citet{Zhang1994} applied this
value to an analysis of the effects of stellar activity on brightness
and concluded that the Sun was $0.2-0.6\%$ dimmer during the Maunder
Minimum.  

\begin{figure}
\plotone{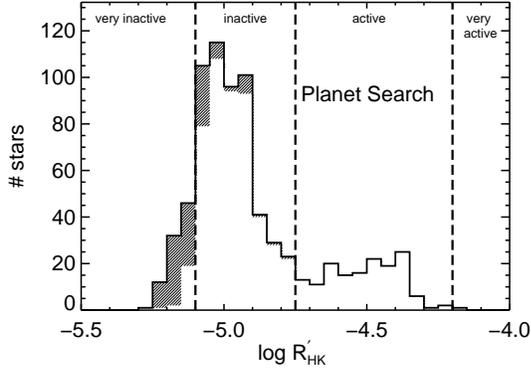}
\caption{Histogram of activity levels of stars in WMBV.  Shaded
  regions are stars for which $\dms > 1$, that is, unambiguously
  evolved stars.  Plotted here are stars with Hipparcos parallaxes
  and $0.55 < \bv <0.9$.  Note that nearly all of the lowest-activity
  stars are subgiants and not main sequence stars in a Maunder
  minimum.  The four activity regions are those of
  \citet{Henry1996}. \label{dist}} 
\end{figure}

\begin{figure}
\plotone{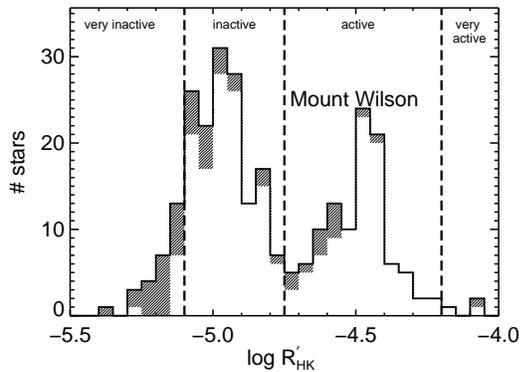}
\caption{As Figure~\ref{dist}, but plotted here are stars with $0.55 <
  \bv < 0.9$ in the Mount Wilson sample.  Again, almost all of the
  ``very inactive'' stars have $\dms > 1$, and so are subgiants and
  not main sequence stars in a Maunder minimum.
\label{mwdist}}  
\end{figure}

\begin{figure}
\plotone{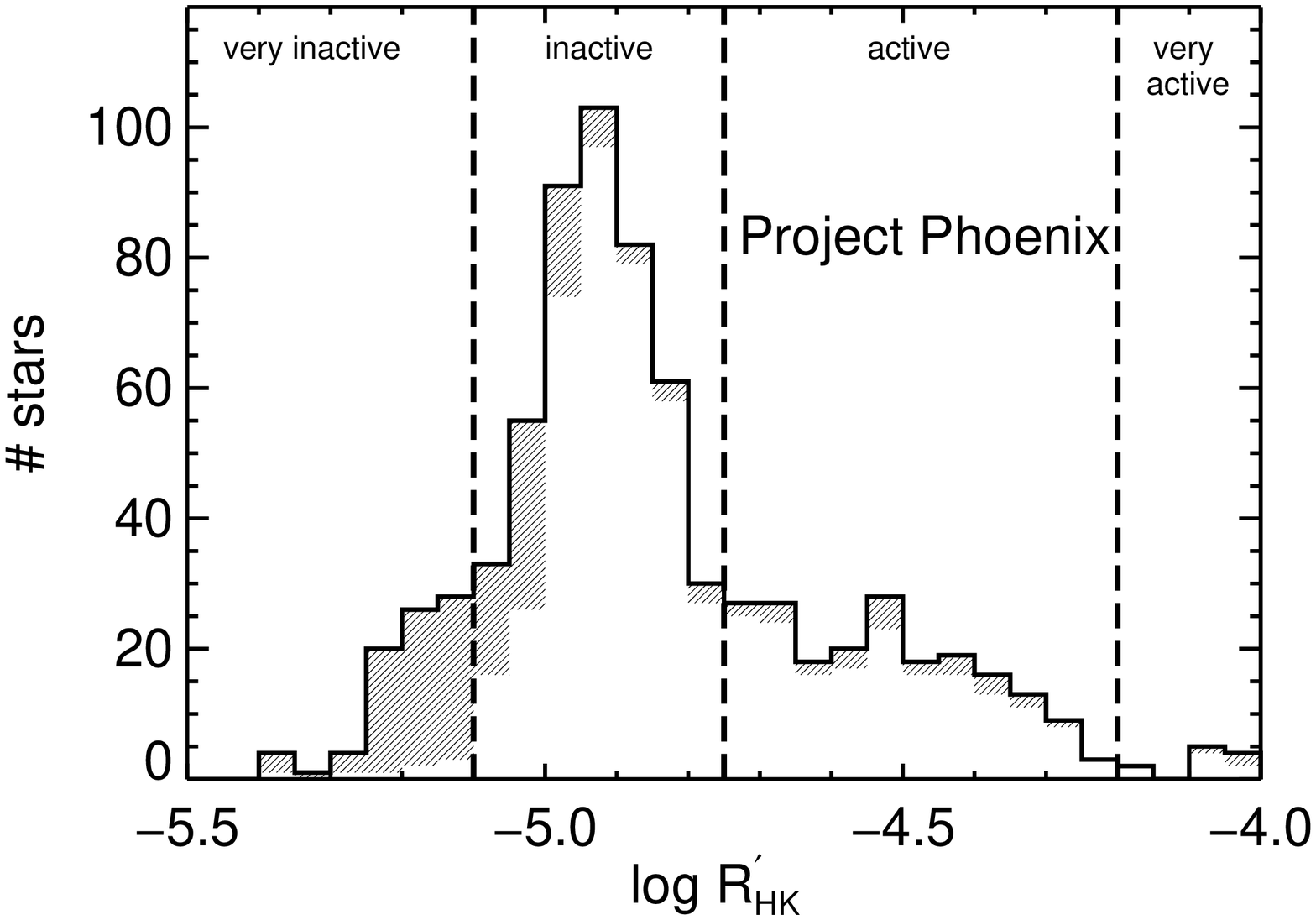}
\caption{As Figure~\ref{dist}, but plotted here are stars in the
  sample of \citet{Henry1996} with $0.55 < \bv < 0.9$.  Again, almost
  all of the ``very inactive'' stars have $\dms\ > 1$, and so are subgiants and
  not main sequence stars in a Maunder minimum.\label{henrydist}}  
\end{figure}

However, most Maunder minimum candidates were identified before the
advent of accurate parallaxes from Hipparcos, making their
evolutionary status difficult to assess accurately.  Here I determine
the evolutionary status of the Maunder minimum stars, showing that
nearly all of these candidates are evolved or subgiant stars, not
sun-like stars in a temporary 
state of very low activity.  Figures~\ref{dist} and \ref{mwdist} show
the distribution of activity levels for stars in the WMBV and
Mount Wilson catalogs.  To derive \dms\ values for the Mount Wilson
data, I applied the same methodology as in \S\ref{ms}, using
only stars with $0.55 < \bv < 0.9$ and $\dms < 4.5$ to
match that of the majority of the stars in WMBV.  The shaded
regions represent stars for which $\dms > 1$, which is a conservative
limit for identifying evolved stars which should exclude all
but the most metal-rich main sequence stars.  Clearly, the ``very
inactive'' portion of the Mount Wilson sample is a population of
evolved stars.  This is also true of the sample of \citet{Henry1996} (see
Figure~\ref{henrydist}).  

\citet{Gray2003} identified nine Maunder minimum candidates with $\lrphk
< -5.1$ from their activity survey, although they qualified this
classification as tentative.  Three of these stars, HD 120066, HD
127334, and HD 195564, are likely subgiants, with $\dms > 0.6$.
Four others, HD 12051, HD 164922, HD57091, and HD 221354, have $\dms >
0.2$ (according to their Hipparcos parallaxes) and $\feh \leq 0$ (from
Table 6 of \citet{Gray2003}) indicating that they are probably at
least slightly evolved.  The final two candidates, HD 65430 and HD
73667, are well observed in the WMBV catalog with \lrphk\ values of
-5.01 and -4.96, respectively, which are at the bottom of the range of
activity levels seen in here for unevolved stars and consistent with
the sun at solar minimum.

\begin{figure}
\plotone{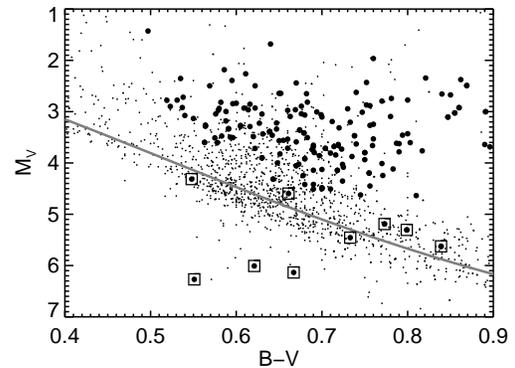}
\caption{Color-magnitude diagram of the stars in the WMBV, Mount Wilson,
  and Project Phoenix surveys, as well as the Maunder minimum candidates
  of \citet{Gray2003}.  Stars with $\lrphk < -5.1$ are plotted with
  larger symbols.  These stars with $\lrphk < -5.1$ have been
  identified in previous works as Maunder minimum candidates;  it is clear that most of these
  stars are, in fact, evolved, and are not sun-like stars in extraordinary
  states of low activity.  The boxed stars have $\dms < 0.4$ and are
  described in the text.  The grey line is the Hipparcos average main
  sequence.\label{mmcm}} 
\end{figure}

Figure~\ref{mmcm} shows all of the stars with $0.4 < \bv < 0.9$ in the
WMBV, Project Phoenix, and Mount Wilson catalogs with Hipparcos
parallaxes, as well as the nine candidates from \citet{Gray2003}, plotted on a
color-magnitude diagram.  Maunder minimum candidates from these works
(those stars with $\lrphk < 5.1$) are plotted with a larger symbol.  In cases
where the WMBV activity measurement differs from that of another
catalog, and the WMBV measurement includes more than two Keck
observations, the WMBV measurement is used.  

Figure~\ref{mmcm} shows that the majority of the stars with $\lrphk <
-5.1$ are actually evolved rather than main sequence stars.
The few Maunder minimum candidates near the main sequence (boxed in
Figure~\ref{mmcm}) are HD 233641 at $\bv = 0.55$, three
of the \citet{Gray2003} Maunder minimum candidates (with $\bv > 0.75$),
and HD186427 at $\bv = 0.66$.  HD 186427 is also on the Lowell
Observatory SSS program, who report $\lrphk \sim -5.02$ and a flat
activity record (J. Hall 2004, private communication;
\citep{Hall2004}).  The inactive star at $\bv=0.73$ just 
below the main sequence is GJ 561, from the Mount Wilson catalog.  The
three low activity stars 
with $\mv > 6$ have very negative \dms-values and so are extremely
metal-poor.  Their low \rphk-values may be due to complications
in extracting reliable activity measurements from such extremely metal-poor
stars, since the shapes of the \caii\ absorption wings can change
dramatically in such objects.

The conclusion that nearly all very inactive stars near the main
sequence are in a state of low activity due to old age is
surprising, since that possibility was considered and rejected by
\citet{Henry1996}, as well as by \citet{Saar1998} in a subsequent
analysis and other authors.  Two of their primary 
arguments are that $a)$ the Sun entered its own Maunder Minimum and
returned to a cyclic state, suggesting that some fraction of sun-like
stars are in an analogous state and $b)$ at least some of the
``Maunder minimum'' stars are young and unevolved (based on rotation
periods and $v\sin{i}$), suggesting that these
low-activity states are not always induced by old age.

There is little doubt of the reality of the Sun's Maunder Minimum
\citep{Eddy1976}, but the question of how to detect an analogous state
in other sun-like stars is difficult.  Since the lowest activity level
observed for unambiguously main sequence stars in Figure~\ref{dmvrphk}
appears to be $\lrphk \sim -5.0$, this may represent the minimum level of
activity in main sequence stars, not the lower value of
\citet{Baliunas1990}.  If so, then low activity alone is not a
sufficient discriminant of a Maunder minimum state.   

The six specific examples of young, chromospherically inactive stars
cited in the above works are: HD 9562 by \citet{Baliunas1995}, HD
167215 and HD 101177 by \citet{Donahue1998}, HD 45067, HD 89744, and
HD 178428 by \citet{Saar1998}.  According to the analysis here
(\S\ref{ms}) five of these stars, HD 9562, HD 167215, HD 45067, HD
89744, and HD 178428 have \dms\ values of 1.5, 0.5, 1.4, 1.1, and 0.8
respectively, suggesting that they are evolved.  That is,
regardless of their age, these stars are not Maunder minimum stars
because they are not on the main sequence.  The \lrphk\ value of HD
101177 in WMBV is -4.95, suggesting that it may not be very inactive,
at all.      

\subsection{The Age-Activity Relation}
\begin{figure}
\plotone{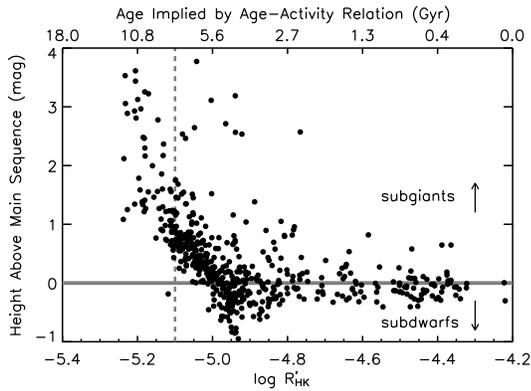}
\caption{\rphk\ vs. \dms\ for WMBV stars with $0.44 < \bv < 0.9$ and
  more than two Keck observations.  The top axis gives the ages
  implied (erroneously) by the age-activity relation.  The vertical
  dashed line at $\lrphk = -5.1$ marks the ostensible threshold for
  Maunder minimum behavior, and the vertical line marks the center of
  the average Hipparcos main sequence. 
  Subdwarfs exhibit a variety of ages and activities, but according to
  the age-activity relation none of them is older than 5.6 Gyr, which
  cannot be correct.  Well above the main sequence there is some
  contamination from metal-rich stars, but the lower envelope of the
  distribution there shows a clear trend of decreasing activity with
  evolution (or perhaps metallicity).  \label{dmvrphk}}   
\end{figure}

Figure~\ref{dmvrphk} shows the activity levels of stars, expressed as \lrphk,
in the WMBV sample as a function of \dms.  This figure includes
only stars for which more than two Keck observations exist and for
which $0.44 < \bv < 0.9$ (the color range where the transformation from $S$
to \rphk\ is calibrated.)  Since \dms\ increases with \feh, the range
of metallicity of the stars in this sample produces a range of
\dms-values for stars of a given age and mass.  On the top axis of
Figure~\ref{dmvrphk} is the age {\it implied} by Equation~\ref{ageact}.

Stars with $\dms \sim 0$ exhibit a variety
of activity levels, as expected for stars with a variety of ages.
Well above the Hipparcos average main sequence ($\dms > 1$), the
bulk of the stars are evolved and exhibit activity levels below the
minimum level of the main sequence stars.  

Since stellar activity decreases roughly as the square root of the age
of a star \citep{Skumanich1972}, it is not surprising that evolved stars will
show low activity levels.  Figure~\ref{dmvrphk} shows,
however, that ages implied by the age-activity relation
(Equation~\ref{ageact}) are incorrect.  In this plot there are
virtually no stars below  
the main sequence with $\lrphk < -5.0$, which, according to
Equation~\ref{ageact}, corresponds to ages $> 5.6$ Gyr.  The WMBV
sample includes many stars with metallicities less than solar, and
since there are no selection criteria in this sample which select
against old main sequence stars some of them must be older than 5.6
Gyr since the Galactic disk is older than that.  Thus, the
age-activity relation must fail for stars with $\lrphk
\lesssim -5.0$.  

One complication to consider is that stars evolve slightly during their main
sequence lifetimes:  according to the isochrones of
\citet{Girardi2002}, even after only 5.6 Gyr a solar mass star will evolve
0.2 magnitude above the ZAMS.  At solar minimum the sun sits at $\dms
= 0, \lrphk=-4.97$, and so as it ages it would not enter the region of
Figure~\ref{dmvrphk} devoid of stars even if the ages implied by
Equation~\ref{ageact} were correct.  

This, however, cannot explain the lack of stars in the region ${\lrphk
  < -5.0, \dms < 0}$ because the WMBV sample is not composed entirely
of solar mass, solar metallicity stars, as Figure~\ref{fefit} makes
clear (for the Sun, 
$\bv = 0.656$).  Stars with $M < M_\odot$ evolve more slowly than more
massive stars, and so will have lower \dms-values than their
solar-type counterparts of the same age.  It is these less-massive stars' apparent
absence from Figure~\ref{dmvrphk}, as well as the apparent absence of
old stars with $\feh < 0$, which proves the breakdown of
Equation~\ref{ageact}.   

\citet{Soderblom1991} conducted an analysis of the age-activity
relation using slightly evolved F stars.  The stars used in this
calibration are somewhat bluer ($\bv < 0.55$) than most of the stars
in Figure~\ref{dmvrphk}, and generally have $\lrphk > -5.0$ , although
  Hipparcos parallaxes confirm that these are indeed evolved stars.   
Since the WMBV sample has few stars with $\bv < 0.55$  this
implies that the relationship between low activity and evolution
demonstrated here may be not be valid for stars with $\bv <
0.55$.  Indeed, \citet{do Nascimento2003} noted that such subgiants
appear to have high \caii\ line core emission independent of their
rotation, unlike subgiants with $\bv > 0.55$. 

\subsection{What is a Maunder minimum Star?}

If we define a Maunder minimum star as a sun-like star which enters,
from a cyclic activity state, a long period of inactivity, then most
of the candidates identified in previous works are not Maunder minimum
stars.  These stars have \rphk\-values levels common to stars
above the main sequence and uncommon to stars on it, so cannot be said
to be ``sun-like'' for this purpose.  Thus, the fact
that they compose roughly 10\% of a stellar sample does not imply that
the sun will spend roughly 10\% of its time in a Maunder Minimum
state.

If, however, the definition of a Maunder minimum state is not restricted to sun-like
stars then the conclusions here weaken.  While it seems likely that
subgiant stars will be in a permanent state of inactivity, some do
show chromospheric emission and some of the Maunder minimum candidate
stars may be only temporarily inactive, possibly meeting this
definition of a Maunder minimum star.

\subsection{The Sources of Low Activity Measurements \label{rphkproblem}}
Figure~\ref{dmvrphk} shows a trend of decreasing activity with
increasing height above
the main sequence, and I have argued that stars with $\lrphk < -5.1$
are evolved.  While this by itself may explain the low activity of
stars for which $\dms > 1$, there remains the possibility that the
stars only sightly above the main sequence ($0 < \dms < 1$) which show
depressed activity ($\lrphk < -5.0$) may simply be metal-rich, and not
evolved at all.  It may seem implausible that metallicity would be the
source of suppressed activity measurements when evolution provides such
a natural explanation, but there is some evidence that this is the case.

\begin{figure}
\plotone{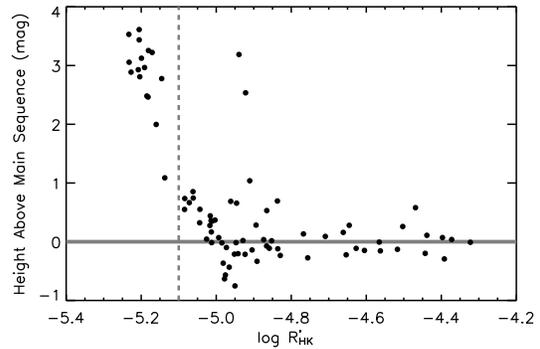}
\caption{As Figure~\ref{dmvrphk}, but restricted to stars with $0.8 <
  \bv < 0.9$.  As Figure~\ref{mmcm} shows, this set of stars includes
  subgiants but very few, if any, slightly evolved stars, as shown by
  the lack of points between $1 < \dms < 2$.  Despite this, for stars
  with $\dms < 1$ and $\lrphk < -5.0$ there
  still appears to be a trend of decreasing activity with increasing
  \dms, just as
  seen in Figure~\ref{dmvrphk}, suggesting that metallicity 
  may be responsible for the decrease. \label{dmvrphk0.8}}
\end{figure}

Figure~\ref{dmvrphk0.8} shows \lrphk\ vs. \dms, as
Figure~\ref{dmvrphk}, but with a sample restricted to stars with $0.8 <
\bv < 0.9$.  These stars, as shown in Figure~\ref{mmcm}, are almost
exclusively subgiants and main-sequence stars with very few, if any,
slightly evolved stars.  This is because no main sequence stars with
these colors in the galactic disk are old enough to have evolved off of the main
sequence.  Figure~\ref{dmvrphk0.8} shows a population of subgiants
with $\dms > 2$ which have very low activity levels.  But
even stars with $\dms < 1$ show a lower envelope of \rphk-values which
decreases with \dms, despite the ostensible lack of significant
evolution in this population.  Since metallicity is presumably the
source of the spread in \dms\ for stars with $\dms < 1$, this suggests
that metallicity may play a role in decreasing \caii\ emission in these
stars.

What is the source of the roughly linear decrease in \rphk\ with
increasing \dms\
on the left side of Figure~\ref{dmvrphk}?  \citet{Rutten1984} noted a
weak gravity dependence for stars in this color range on the minimum
\caii\ flux, which is consistent with the effect seen here.  \citet{Schrijver1989} note that the
minimum \caii\ flux level has several sources:  in addition to a 
basal chromospheric flux there is emission from the stellar
photosphere in the line wings and core.  The flux contribution from
either of these components could be a function of gravity or
metallicity.  \rphk\ ostensibly does not include any photospheric
flux, since that component is subtracted out.  But the methods used
for this removal of the
photospheric component by the catalogs used here did not include any
corrections for gravity or metallicity.  This means that any
dependence in the photospheric flux on these quantities remain in
the final \rphk-value.

\section{Future Work}

\citet{Baliunas1995} and \citet{Soderblom1991} revealed a wealth of
information about the chromospheric behavior of late-type stars as a
function of age.  Now that the Hipparcos program has determined
accurate parallaxes for many of these stars, these studies can be
refined with evolution and gravity as additional parameters.  For instance,
\citet{Soderblom1991} included slightly evolved F dwarfs in their
calibration of the age-activity relation and then applied the relation
to the Mount Wilson survey data to determine the star formation
history of nearby stars.  This study needs to be expanded to account
for evolution and to include lower-mass, evolved stars.

In addition, accurate and homogeneous metallicities for stars (with careful
attention paid to evolution and the effects of gravity on metallicity
measurements) will allow for more precise age measurements and a
more accurate age-activity relationship.  A forthcoming work will
present such abundance measurements for the Planet Search stars
\citep{Valenti2004}. 

HD 233641, HD 186427, GJ 561, and other very inactive stars need to be
studied further to 
determine their metallicity, evolutionary, and activity status with more confidence.
Establishing a firm lower boundary to the range of activities of
unevolved stars will help to determine whether Maunder minima
are extremely rare or are simply not indicated by low activity
alone. \citet{Saar1998} appropriately uses the term ``flat-activity'' to 
refer to stars which may be in a Maunder minimum, pointing out that
a lack of activity variations may be an indicator of a Maunder minimum state. 

\section{Conclusions}

As has been noted elsewhere \citep{do Nascimento2003}, evolved stars with
$0.55 < \bv < 0.9$ have lower levels of activity than main sequence stars.
Slightly evolved stars, even those just a fraction of a magnitude
above the main sequence, are no exception, and have been mistaken for
sun-like stars in extraordinarily low states of activity.  This has led to
speculation that these stars are in states analogous to the Maunder
Minimum of the Sun in the 17th century.  

In fact, unevolved stars almost never have activity levels such that
$\lrphk \lesssim -5.1$.  This implies that if other stars do undergo
Maunder minima of their own, then it is either a rare occurrence
undetected in activity surveys or it is not necessarily indicated by
the condition $\lrphk \lesssim -5.1$.  Other tests, such as looking for a lack
of activity variations or highly discrepant activity behavior among
binary components, may prove more useful.  To date, there is no
unambiguous identification of another star in a Maunder minimum state.

The canonical calculation used to derive \rphk\ from $S$-values may need
to be revised to account for variations in gravity and 
metallicity when subtracting the photospheric component of emission in
the \caii\ line cores.  Further, the age-activity relation derived
from these \rphk\ measurements and commonly used to derive stellar
ages fails for stars with $\lrphk < -5.0$, and needs to be revisited.   

\acknowledgments
I thank Geoff Marcy, John Johnson, Debra Fischer, and Steve Saar for
their input 
in the writing of this paper.  I am also indebted to Gibor Basri for
illuminating discussions regarding the nature of low-\rphk\ stars.

I also wish to recognize and acknowledge the
very significant cultural role and reverence that the summit of Mauna
Kea has always had within the indigenous Hawaiian community.  We are
most fortunate to have the opportunity to conduct observations from
this mountain.

This research has made use of the SIMBAD database, operated at CDS,
Strasbourg, France, and of NASA's Astrophysics Data System
Bibliographic Services, and is made possible by the generous support
of Sun Microsystems, the NSF, and NASA through grant NAG5-12090.

\end{document}